\documentclass[a4paper]{spie}  

\usepackage[]{graphicx}
\usepackage{url}
\usepackage{ifpdf}
\usepackage{hyperref}
\hypersetup{breaklinks,colorlinks,citecolor=blue}

\usepackage{amssymb}
\usepackage{amsmath}
\usepackage{mathtools}
\DeclarePairedDelimiter{\ceil}{\lceil}{\rceil}
\DeclarePairedDelimiter{\floor}{\lfloor}{\rfloor}
\usepackage{subfigure}
\usepackage{color}
\usepackage{subfigure}
\usepackage{geometry}
\newcommand{\eqn}[1]{(#1)}

\newcommand{\fig}[1]{Fig.~#1}

\newcommand{\sectn}[1]{Sec.~#1}

\newcommand{\etal}{\mbox{\it et al.}}
\newcommand{\eg}{\mbox{\it e.g.}}
\newcommand{\ie}{\mbox{\it i.e.}}

\newcommand{\cmb}{{CMB}}
\newcommand{\cmbtext}{{cosmic microwave background}}

\newcommand{\healpix}{{\tt HEALPix}}

\newcommand{\spcend}{\ensuremath{\:}}

\newcommand{\cconj}{\ensuremath{\ast}} 
 
\newcommand{\reals}{\ensuremath{\mathbb{R}}}

\newcommand{\realsnz}{\ensuremath{\mathbb{R}^{+}_{\ast}}}
\newcommand{\integers}{\ensuremath{\mathbb{Z}}}
\newcommand{\naturals}{\ensuremath{\mathbb{N}}}

\newcommand{\ltwo}{\ensuremath{\mathrm{L}^2}}
\newcommand{\sphere}{\ensuremath{{\mathbb{S}^2}}}
\newcommand{\sothree}{\ensuremath{{\mathrm{SO}(3)}}}

\newcommand{\dx}{\ensuremath{\mathrm{\,d}}}
\newcommand{\dmu}[1]{\ensuremath{\dx \Omega(#1)}}

\newcommand{\deul}[1]{\ensuremath{\dx \varrho(#1)}}

\newcommand{\innerp}[2]{\ensuremath{\langle {#1},\: {#2} \rangle}}

\newcommand{\sa}{\ensuremath{\omega}}
\newcommand{\saa}{\ensuremath{\theta}}
\newcommand{\sab}{\ensuremath{\varphi}}
\newcommand{\sas}{\ensuremath{\saa, \sab}}
\newcommand{\eul}{\ensuremath{\mathbf{\rho}}}
\newcommand{\euls}{\ensuremath{\eula, \eulb, \eulc}}
\newcommand{\eula}{\ensuremath{\alpha}}
\newcommand{\eulb}{\ensuremath{\beta}}
\newcommand{\eulc}{\ensuremath{\gamma}}
\newcommand{\eulai}{\ensuremath{a}}
\newcommand{\eulbi}{\ensuremath{b}}
\newcommand{\eulci}{\ensuremath{g}}
\newcommand{\eulaiang}{\ensuremath{\eula_\eulai}}
\newcommand{\eulbiang}{\ensuremath{\eulb_\eulbi}}
\newcommand{\eulciang}{\ensuremath{\eulc_\eulci}}
\newcommand{\el}{\ensuremath{\ell}}
\newcommand{\m}{\ensuremath{m}}
\newcommand{\n}{\ensuremath{n}}

\newcommand{\elmax}{\ensuremath{{L}}}

\newcommand{\nmax}{\ensuremath{{N}}}

\newcommand{\p}{\ensuremath{^\prime}}

\newcommand{\kron}[2]{\ensuremath{\delta_{{#1}{#2}}}}

\newcommand{\shfarg}[3]{\ensuremath{Y_{#1#2}({#3})}}

\newcommand{\shf}[2]{\ensuremath{Y_{#1#2}}}

\newcommand{\shc}[3]{\ensuremath{{#1}_{{#2}{#3}}}}
\newcommand{\shcc}[3]{\ensuremath{{#1}_{{#2}{#3}}^\cconj}}

\newcommand{\dmatbig}{\ensuremath{D}}
\newcommand{\Dlmn}{\ensuremath{ \dmatbig_{\m\n}^{\el} }}
\newcommand{\Dlmnc}{\ensuremath{ \dmatbig_{\m\n}^{\el\cconj} }}

\newcommand{\Dlmnp}{\ensuremath{ \dmatbig_{\m\n}^{\el}(\eul) }}
\newcommand{\Dlmnpc}{\ensuremath{ \dmatbig_{\m\n}^{\el\cconj}(\eul) }}

\newcommand{\dmatsmall}{\ensuremath{d}}
\newcommand{\dlmn}{\ensuremath{ \dmatsmall_{\m\n}^{\el} }}

\newcommand{\dlmnhalfpi}[3]{\ensuremath{ \Delta_{{#2}{#3}}^{#1} }}

\newcommand{\wigc}[4]{\ensuremath{{#1}^{#2}_{{#3}{#4}}}}

\newcommand{\rot}{\ensuremath{\mathcal{R}}}
\newcommand{\rotarg}[1]{\ensuremath{\mathcal{R_{#1}}}}

\newcommand{\rotmatarg}[1]{\ensuremath{\mathbf{R}_{#1}}}

\newcommand{\f}{\ensuremath{f}}

\newcommand{\wav}{\ensuremath{\psi}}

\newcommand{\wavs}{\ensuremath{\Phi}}
\newcommand{\wcoeff}{\ensuremath{W}}
\newcommand{\scoeff}{\ensuremath{W}}
\newcommand{\wscale}{\ensuremath{j}}
\newcommand{\wscalemax}{\ensuremath{J}}

\newcommand{\dilparam}{\ensuremath{\alpha}}
\newcommand{\wavker}{\ensuremath{\kappa}}
\newcommand{\wavsteer}{\ensuremath{s}}
\newcommand{\steerinterp}{\ensuremath{z}}

\newcommand{\sumlmn}{\ensuremath{\sum_{\el=0}^{\infty} \sum_{\m=-\el}^\el} \sum_{\n=-\el}^\el}

\newcommand{\summ}{\ensuremath{\sum_{\m=-\el}^\el}}

\newcommand{\sumulmn}{\ensuremath{\sum_{\el\m\n}}}

\newcommand{\saai}{\ensuremath{t}}
\newcommand{\sabi}{\ensuremath{p}}
\newcommand{\saaiang}{\ensuremath{\saa_\saai}}
\newcommand{\sabiang}{\ensuremath{\sab_\sabi}}

\newcommand{\qweight}{\ensuremath{q}}

\newcommand{\order}{\ensuremath{\mathcal{O}}}

\newcommand{\conv}{\ensuremath{\star}}

\renewcommand{\eqn}[1]{Eqn.~(#1)}

\renewcommand{\wav}{\ensuremath{\Psi}}

\title{On the computation of directional scale-discretized\\wavelet transforms on the sphere}

\author{Jason D. McEwen\supit{a,b}, Pierre Vandergheynst\supit{c} and Yves~Wiaux\supit{c,d,e}
\skiplinehalf
\supit{a} Department of Physics and Astronomy, University College London (UCL), \\
London WC1E 6BT, UK\\
\supit{b} Mullard Space Science Laboratory (MSSL), University College London (UCL), \\
Surrey RH5 6NT, UK\\
\supit{c} Institute of Electrical Engineering, Ecole Polytechnique F{\'e}d{\'e}rale de Lausanne (EPFL), \\ Lausanne 1015, Switzerland; \\
\supit{d} Department of Medical Radiology, University Hospital Center (CHUV) and University of Lausanne (UNIL), CH-1011 Lausanne, Switzerland; \\ 
\supit{e} Department of Radiology and Medical Informatics, University of Geneva (UniGE), \\ Geneva 1211,
  Switzerland
}

\authorinfo{Further author information:\\JDM: E-mail: \href{mailto:jason.mcewen@ucl.ac.uk}{jason.mcewen@ucl.ac.uk}, URL: \href{http://www.jasonmcewen.org/}{http://www.jasonmcewen.org/}}

\begin{document} 
\maketitle 


\begin{abstract}
  We review scale-discretized wavelets on the sphere, which are
  directional and allow one to probe oriented structure in data
  defined on the sphere.  Furthermore, scale-discretized wavelets
  allow in practice the exact synthesis of a signal from its wavelet
  coefficients.  We present exact and efficient algorithms to compute
  the scale-discretized wavelet transform of band-limited signals on
  the sphere.  These algorithms are implemented in the publicly
  available {\tt S2DW} code.  We release a new version of {\tt S2DW}
  that is parallelized and contains additional code optimizations.
  Note that scale-discretized wavelets can be viewed as a directional
  generalization of needlets.  Finally, we outline future improvements
  to the algorithms presented, which can be achieved by exploiting a
  new sampling theorem on the sphere developed recently by some of the
  authors.
\end{abstract}


\section{INTRODUCTION} 

Wavelets on the sphere have found widespread application in fields
such as cosmology (\eg\ Ref.~\citenum{planck2013-p09}) and geophysics
(\eg\ Ref.~\citenum{simons:2011}), where data are observed on a
spherical domain.  The ability of wavelets to probe spatially
localized, scale-dependent features in signals has become an
instrumental tool to relate data to physical theories, as many
physical processes are spatially localised but manifest on particular
physical scales.  Since such data-sets are of considerable size, often
containing tens of millions of pixels, efficient algorithms are also
of central importance.

Many wavelet transforms on the sphere have been
developed.\cite{antoine:1998, antoine:1999, baldi:2006,
  marinucci:2008, leistedt:s2let_axisym, mcewen:2006:cswt2,
  mcewen:2006:fcswt, narcowich:2006, starck:2006, wiaux:2005,
  wiaux:2005c, wiaux:2006:review, yeo:2008}. Of particular note are
so-called needlets, which have been applied extensively to the
analysis of observations of the \cmbtext\ (\cmb).  Needlets have found
broad application since they exhibit many useful properties and the
needlet transform can be computed relatively straightforwardly.
However, needlets are axisymmetric wavelets and thus cannot be used to
probe directional or oriented structure in data on the sphere.

In this article we review the scale-discretized wavelets on the sphere
developed by Wiaux \etal\cite{wiaux:2007:sdw}, which were developed
independently of needlets, about the same time, and which share many
of the useful properties of needlets.  However, scale-discretized
wavelets are also directional, allowing one to probe oriented
structure in data.  Directional wavelets are very useful for the
analysis of signals on the sphere with oriented structure; for
example, the \cmb\ fluctuations induced by cosmic
strings,\cite{hammond:2009,planck2013-p20} a well-motivated but as yet
unobserved phenomenon.

In addition to reviewing scale-discretized wavelets on the sphere, we
present exact and efficient algorithms to compute the corresponding
wavelet analysis and synthesis, \ie\ the forward and inverse wavelet
transform respectively.  These algorithms are implemented in the
publicly available {\tt S2DW}\footnote{\url{http://www.s2dw.org/}}
code.  Although the {\tt S2DW} package has been available publicly for
some time, these algorithms have not yet been described in the
literature.  Furthermore, we release a new version (1.1) of {\tt S2DW}
that is parallelized and contains additional code optimizations.

The remainder of this article is organized as follows.  In
\sectn{\ref{sec:transform}} we review the scale-discretized wavelet
transform on the sphere.  In \sectn{\ref{sec:computation}} we present
our algorithms to compute the wavelet transform exactly and
efficiently.  Numerical experiments are also performed to evaluate the
speed and accuracy of the algorithms.  Concluding remarks are made in
\sectn{\ref{sec:summary}}, where we also outline future optimizations
of the algorithms by exploiting recent developments\cite{mcewen:fssht}
made by some of the authors pertaining to sampling theorems on the
sphere.

\section{DIRECTIONAL SCALE-DISCRETIZED WAVELETS} 
\label{sec:transform}

The directional scale-discretized wavelet transform supports the
analysis of oriented spatially localised, scale-dependent features in
signals on the sphere.  In this section we review the
scale-discretized wavelet framework on the sphere developed by Wiaux
\etal\cite{wiaux:2007:sdw}.  Firstly, we review the wavelet transform
itself, before secondly reviewing the construction of admissible
scale-discretized wavelets.

\subsection{Wavelet transform}

The scale-discretized wavelet transform of a function $\f \in
\ltwo(\sphere)$ on the sphere \sphere\ is defined by the directional
convolution of \f\ with the wavelet $\wav^\wscale \in
\ltwo(\sphere)$.  The wavelet coefficients $\wcoeff^{\wav^\wscale} \in
\ltwo(\sothree)$ thus read
\begin{equation}
  \label{eqn:analysis}
  \wcoeff^{\wav^\wscale}(\eul) \equiv ( \f \conv \wav^\wscale) (\eul)
  = \innerp{\f}{\rotarg{\eul}\wav^\wscale}
  = \int_\sphere \dmu{\sa} \f(\sa) (\rotarg{\eul}\wav^\wscale)^\cconj(\sa)
  \spcend ,
\end{equation}
where $\sa=(\sas) \in \sphere$ denotes spherical coordinates with
colatitude $\saa \in [0,\pi]$ and longitude $\sab \in [0,2\pi)$,
$\dmu{\sa} = \sin\saa \dx\saa \dx\sab$ is the usual rotation
invariant measure on the sphere, and $\cdot^\cconj$ denotes complex
conjugation.  The rotation operator is defined by 
\begin{equation}
  (\rotarg{\eul} \wav^\wscale) \equiv \wav^\wscale(\rotmatarg{\eul}^{-1} \cdot \sa)
  \spcend ,
\end{equation}
where $\rotmatarg{\eul}$ is the three-dimensional rotation matrix
corresponding to $\rotarg{\eul}$.  Rotations are specified by elements of
the rotation group $\sothree$, parameterized by the Euler angles
$\eul=(\euls) \in \sothree$, with $\eula \in [0,2\pi)$, $\eulb \in
[0,\pi]$ and $\eulc \in [0,2\pi)$.  We adopt the $zyz$ Euler
convention corresponding to the rotation of a physical body in a
\emph{fixed} coordinate system about the $z$, $y$ and $z$ axes by
$\eulc$, $\eulb$ and $\eula$, respectively.  The wavelet transform
\eqn{\ref{eqn:analysis}} thus probes directional structure in the
signal of interest \f, where \eulc\ corresponds to the orientation
about each point on the sphere $(\sas) = (\eulb, \eula)$.

The wavelet scale $\wscale \in \naturals_0$ encodes the localization
extent of $\wav^\wscale$, which is closely related to the wavelet
construction and is discussed in
\sectn{\ref{sec:transform:construction}}.  For now it is sufficient to
note that the wavelet scales $\wscale$ are discrete (hence the name
scale-discretized wavelets), which affords the exact synthesis of a
function from its wavelet (and scaling) coefficients.

The wavelet coefficients encode only the high-frequency, detail
information contained in the signal \f; scaling coefficients must be
introduced to represent the low-frequency, approximation information of the
signal.  The scaling coefficients $\scoeff^\wavs \in \ltwo(\sphere) $
are given by the convolution of \f\ with the axisymmetric scaling
function $\wavs \in \ltwo(\sphere)$ and read
\begin{equation}
  \label{eqn:analysis_scaling}
  \wcoeff^{\wavs}(\sa) \equiv ( \f \conv \wavs) (\sa)
  = \innerp{\f}{\rotarg{\sa}\wavs}
  = \int_\sphere \dmu{\sa\p} \f(\sa\p) (\rotarg{\sa}\wavs)^\cconj(\sa\p)
  \spcend ,
\end{equation}
where $\rotarg{\sa} = \rot_{(\sab,\saa, 0)}$.  Note that the scaling
coefficients live on the sphere, and not the rotation group \sothree,
since we do not probe the directional structure of the
low-frequency, approximation information of \f.  An extension to a directional
scaling function could be made trivially, however it is not of significant
practical use.  Typically, a signal of interest is decomposed into its
scaling and wavelet coefficients and the wavelet coefficients are
analyzed or manipulated, while the scaling coefficients remain
untouched.  

Provided an admissibility condition holds, the signal \f\ can be
synthesised perfectly from its wavelet and scaling coefficients by
\begin{equation}
  \label{eqn:synthesis}
  \f(\sa) = 2\pi \int_\sphere \dmu{\sa\p} 
  \scoeff^\wavs(\sa\p) (\rotarg{\sa\p} L^{\rm d} \wavs)(\sa)
  +
  \sum_{\wscale=0}^\wscalemax \int_\sothree \deul{\eul}
  \wcoeff^{\wav^\wscale}(\eul) (\rotarg{\eul} L^{\rm d} \wav^\wscale)(\sa)
  \spcend ,
\end{equation}
where $\deul{\eul} =  \sin\eulb \dx\eula \dx\eulb \dx\eulc$ is the
usual invariant measure on \sothree\ and $\wscalemax$ is the maximum analysis
depth considered, \ie\ $0 \leq \wscale \leq \wscalemax$ (discussed in more detail in
\sectn{\ref{sec:transform:construction}}). The operator $L^{\rm d}$ is
defined by its action on the harmonic coefficients of functions
$h\in\ltwo(\sphere)$:
\begin{equation}
   \shc{(L^{\rm d}h)}{\el}{\m} \equiv \frac{2\el+1}{8\pi^2}
   \shc{h}{\el}{\m} 
   \spcend .
\end{equation}
where $\shc{h}{\el}{\m} = \innerp{h}{\shf{\el}{\m}}$ are the spherical
harmonic coefficients of $h$ and $\shf{\el}{\m} \in \ltwo(\sphere)$
are the spherical harmonics\cite{varshalovich:1989} with $\el \in
\naturals_0$ and $\m \in \integers$, such that $\vert \m \vert \leq
\el$.

The admissibility condition under which a function \f\ can be
synthesised perfectly from its wavelet and scaling coefficients is
given by the following resolution of the identity:
\begin{equation}
  \label{eqn:admissibility}
  \vert \shc{\wavs}{\el}{0} \vert^2 + \sum_{\wscale=0}^\wscalemax
  \summ \vert \shc{\wav}{\el}{\m}^\wscale\vert^2 = 1 
  \spcend ,  
  \quad \forall\el 
  \spcend ,
\end{equation}
where $\shc{\wavs}{\el}{0} \kron{\m}{0} =
\innerp{\wavs}{\shf{\el}{\m}}$ and $\shc{\wav}{\el}{\m}^\wscale =
\innerp{\wav}{\shf{\el}{\m}}$ are the spherical harmonic coefficients
of $\wavs$ and $\wav^\wscale$, respectively.  Typically, we consider
band-limited functions, \ie\ functions such that $\shc{\f}{\el}{\m} =
0$, $\forall \el \geq \elmax$, where $\shc{\f}{\el}{\m} =
\innerp{\f}{\shf{\el}{\m}}$, for which case wavelet analysis and
synthesis can be computed exactly in practice (see
\sectn{\ref{sec:computation}}).

\subsection{Wavelet construction} 
\label{sec:transform:construction}

The scale-discretized wavelets are constructed in such a way to ensure
the admissibility criterion \eqn{\ref{eqn:admissibility}} is
satisfied.  Furthermore, the wavelets are defined in harmonic space in
the factorized form:
\begin{equation}
  \label{eqn:wav_factorized}
  \shc{\wav}{\el}{\m}^\wscale \equiv \wavker^\wscale(\el) \shc{\wavsteer}{\el}{\m}
  \spcend,
\end{equation}
in order to essentially control their angular and directional
localization separately, through the kernel $\wavker^\wscale$ and
directionality component, with harmonic coefficients
$\shc{\wavsteer}{\el}{\m}$, respectively; we discuss each in turn.

Without loss of generality, the directionality component $\wavsteer
\in \ltwo(\sphere)$, with harmonic coefficients
$\shc{\wavsteer}{\el}{\m} = \innerp{\wavsteer}{\shf{\el}{\m}}$, is
defined to impose
\begin{equation}
	\sum_{\vert \m \vert \leq \el} \vert \shc{\wavsteer}{\el}{\m} \vert^2 = 1
	\spcend, 
\end{equation}
for all values of \el\ for which $\shc{\wavsteer}{\el}{\m}$ are
non-zero for at least one value of \m.  The angular localisation
properties of the wavelet $\wav^j$ are then controlled largely by the
kernel and to a lesser extent by the directional component, while the
directional component controls precisely the directional properties of
the wavelet (\ie\ the behaviour of the wavelet with respect to the
azimuthal variable \sab, when centered on the North pole).
Furthermore, we impose the zero-mean condition
$\shc{\wavsteer}{0}{0}=0$ and an azimuthal band-limit \nmax\ on the
directional component such that $\shc{\wavsteer}{\el}{\m}=0$, $\forall
\el,\m$ with $\vert \m \vert \geq \nmax$.

As soon as one imposes an azimuthal band-limit, one recovers steerable
wavelets\cite{wiaux:2005c, wiaux:2007:sdw,
  wiaux:2010:wavelets_sphere}, where the wavelet rotated about itself
(\ie\ rotated about the $z$ axis when centered on the North pole) can
be expressed as a finite weighted linear combination of wavelets
specified at base orientations. This property is expressed
equivalently on the directional component \wavsteer\ by
\begin{equation}
  \label{eqn:steerability_wav}
  \wavsteer_\eulc(\sa) = 
  \sum_{\eulci=0}^{M-1} \steerinterp(\eulc - \eulciang) s_{\eulciang}(\sa)
  \spcend,
\end{equation}
where $\wavsteer_\eulc \equiv \rot_{(0,0,\eulc)} \wavsteer$. The
rotation angles $\eulciang$ are equispaced and are defined explicitly
in \sectn{\ref{sec:computation:quadrature}}.  We consider wavelets
with even or odd azimuthal symmetry (\ie\ even or odd symmetry when
rotated about itself by $\pi$, when centered on the North pole), in
which case $M=N$.  The interpolating function $\steerinterp(\gamma)$ is
independent of $\wavsteer$ and is defined by its Fourier coefficients
$\steerinterp_\n$ which take particularly simple forms for wavelets with even/odd
azimuthal symmetry (see Ref.~\citenum{wiaux:2007:sdw} for further
detail).  Due to the linearity of the wavelet transform, the
steerability property is transferred to the wavelet coefficients
themselves, yielding
\begin{equation}
  \label{eqn:steerability_wcoeff}
  \wcoeff^{\wav^\wscale}(\euls) = 
  \sum_{\eulci=0}^{M-1} \steerinterp(\eulc - \eulciang) 
  \wcoeff^{\wav^\wscale}(\eula,\eulb,\eulciang)  
  \spcend.
\end{equation}
Steerability is thus a very useful property for practical
applications; for examples of steerable wavelet analyses on the sphere
see Refs.~\citenum{mcewen:2007:isw2, vielva:2006}. 

Following Ref.~\citenum{wiaux:2007:sdw} we define the directionality
coefficients to satisfy the criteria just discussed but also to ensure
a convenient form for the directional auto-correlation function of the
wavelet, yielding
\begin{equation}
	\shc{\wavsteer}{\el}{\m}
	= \eta_{\nmax}\beta_{(\nmax,\m)}
	\sqrt{\frac{1}{2^{\gamma_{(\nmax,\el)}}}
	\Biggl({\gamma_{(\nmax,\el)}\atop \frac{\gamma_{(\nmax,\el)}-\m}{2}}\Biggr)	
	}
	\spcend,
\end{equation}
with $\eta_{\nmax}=1$ for even values of $\nmax-1$,
$\eta_{\nmax}=i=\sqrt{-1}$ for odd values of $\nmax-1$,
$\beta_{(\nmax,m)}=\bigl(1-(-1)^{\nmax+m}\bigr)/2$, and
$\gamma_{(\nmax,\el)}=\min
\bigl\{\nmax-1,\el-\bigl(1+(-1)^{\nmax+\el}\bigr)/2 \bigr\}$.  With
this definition one also recovers a wavelet with odd (even) azimuthal
symmetry for \nmax\ even (odd).

\begin{figure}
\centering
\includegraphics[width=0.75\textwidth]{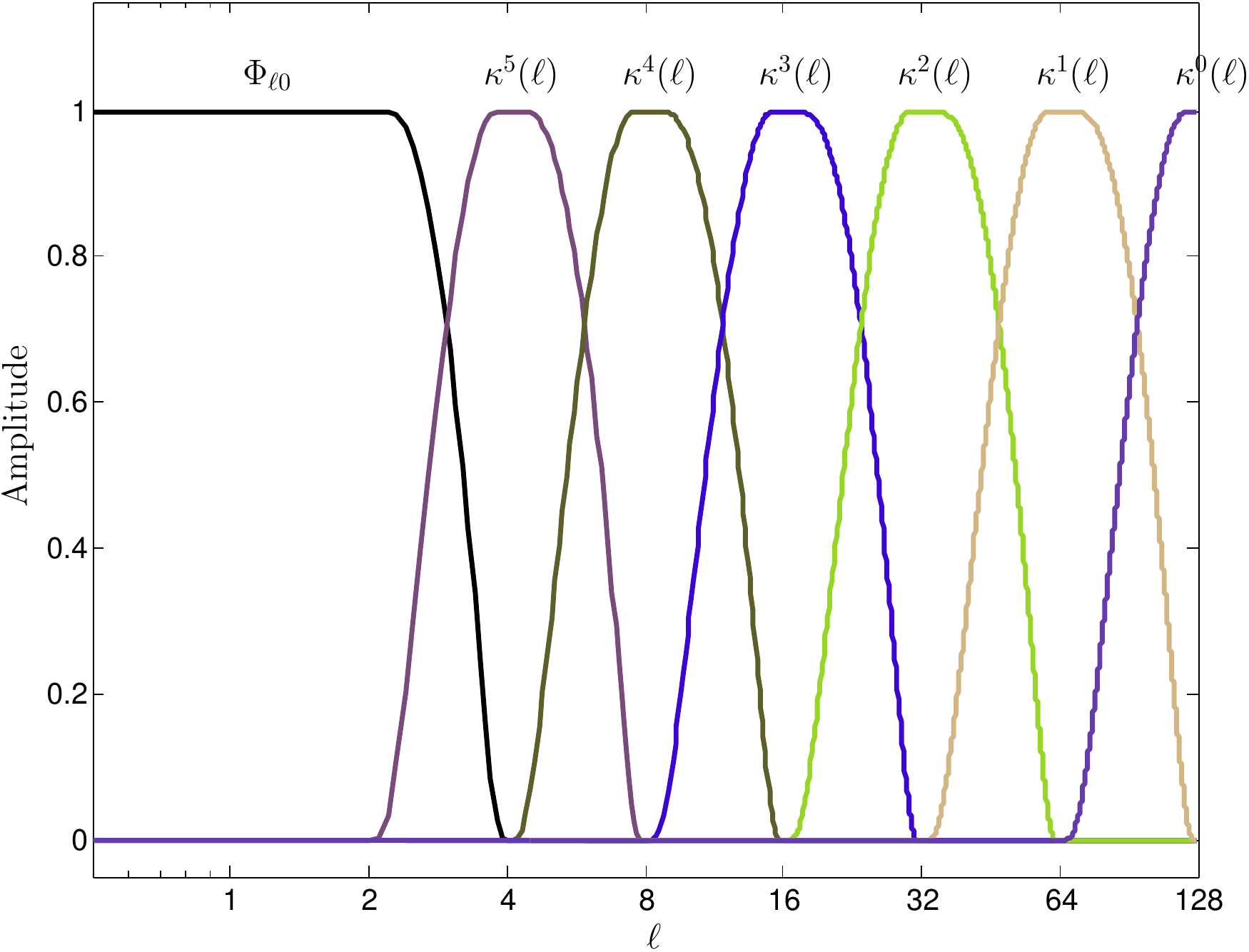}
\caption{Scale-discretized wavelet tiling in harmonic space
  ($\elmax=128$, $\nmax=3$, $\wscalemax=5$, $\dilparam=2$).} 
\label{fig:tiling}
\end{figure}

The kernel $\wavker^\wscale(t)$ is a positive real function, with
argument $t \in \reals$, although $\wavker(t)$ is only evaluated for
natural arguments $t=\el$ in \eqn{\ref{eqn:wav_factorized}}.  The
kernel controls the angular localization of the wavelet and is
constructed to be a smooth function with compact support, as follows.
Consider the infinitely differentiable Schwartz function with compact
support $t \in [\dilparam^{-1}, 1]$, for dilation parameter $\dilparam
\in \realsnz$, $\dilparam>1$:
\begin{equation}
  \wavsteer_\dilparam(t) 
  \equiv \wavsteer\biggl( \frac{2\dilparam}{\dilparam-1} (t-\dilparam^{-1})-1\biggr)
  \spcend,
  \textrm{\quad with \quad } s(t) \equiv \Biggl\{ \begin{array}{ll} 
  \ 
  {\rm exp}\bigl(-(1-t^2)^{-1}\bigr), & t\in[-1,1] \\ \  0, & t \notin [-1,1]\end{array} \spcend.
\end{equation}
We then define the smoothly decreasing function $k_\dilparam$ by
\begin{equation}
  k_\dilparam(t) \equiv \frac{\int_{t}^1\frac{{\rm d}t^\prime}{t^\prime}s_\dilparam^2(t^\prime)}{\int_{\dilparam^{-1}}^1\frac{{\rm d}t^\prime}{t^\prime}s_\dilparam^2(t^\prime)}, 
\end{equation}
which is unity for $t<\dilparam^{-1}$, zero for $t>1$, and is smoothly
decreasing from unity to zero for $t \in [\dilparam^{-1},1]$.  We
define the wavelet kernel generating function by
\begin{equation}
  \wavker_\dilparam(t) \equiv \sqrt{ k_\dilparam(\dilparam^{-1} t) - k_\dilparam(t) }
  \spcend,
\end{equation}
which has compact support $t \in [\dilparam^{-1}, \dilparam]$ and
reaches a peak of unity at $t=1$.  The scale-discretized wavelet
kernel for scale \wscale\ is then defined by
\begin{equation}
  \wavker^\wscale(\el) \equiv  \wavker_\dilparam(\dilparam^{\wscale} \elmax^{-1} \el)
  \spcend,
\end{equation}
which has compact support on $\el \in
\bigl[\floor{\dilparam^{-(1+\wscale)} \elmax},
\ceil{\dilparam^{1-\wscale} \elmax} \bigr]$, where $\floor{\cdot}$ and
$\ceil{\cdot}$ are the floor and ceiling functions respectively, and
reaches a peak of unity at $\dilparam^{-\wscale}\elmax$.\footnote{We
  adopt the same definition as Ref.~\citenum{wiaux:2007:sdw} for the
  wavelet scales \wscale, with increasing \wscale\ corresponding to
  larger angular scales, \ie\ lower frequency content.  Note that this
  differs to the definition adopted in
  Ref.~\citenum{leistedt:s2let_axisym} where increasing \wscale\
  corresponds to smaller angular scales but higher frequency content.}
With this construction the kernel functions tile the harmonic line,
as illustrated in \fig{\ref{fig:tiling}}.

The maximum possible wavelet scale $\wscalemax_\elmax(\dilparam)$ is
given by the lowest integer $\wscale$ for which the kernel peak occurs
at or below $\el=1$, \ie\ by the lowest integer value such that
$\dilparam^{-\wscalemax_\elmax(\dilparam)} \elmax \leq 1$, yielding
$\wscalemax_\elmax(\dilparam) = \ceil{\log_\dilparam(\elmax)}$.  All
wavelets for $\wscale > \wscalemax_\elmax(\dilparam)$ would be
identically null as their kernel would have compact support in $\el
\in (0, 1)$.  The maximum scale to be probed by the wavelets
$\wscalemax$ can be chosen within the range
$0\leq\wscalemax\leq\wscalemax_\elmax(\dilparam)$.  For
$\wscalemax=\wscalemax_\elmax(\dilparam)$ the wavelets probe the
entire frequency content of the signal of interest \f\ except its
mean, encoded in $\shc{\f}{0}{0}$.

To represent the signal content not probed by the wavelets the scaling
function \wavs\ is required, as discussed previously.  Recall that the
scaling function $\wavs$ is chosen to be axisymmetric; hence, we
define the harmonic coefficients of the scaling function by
\begin{equation}
  \shc{\wavs}{\el}{0} 
  \equiv \sqrt{ k_\dilparam(\dilparam^{\wscalemax} \elmax^{-1} \el)} 
  \spcend,
\end{equation}
in order to ensure the scaling function probes the signal content not
probed by the wavelets.

\begin{figure}
\centering
\subfigure[$\wscale=0$]{\includegraphics[width=0.28\textwidth]{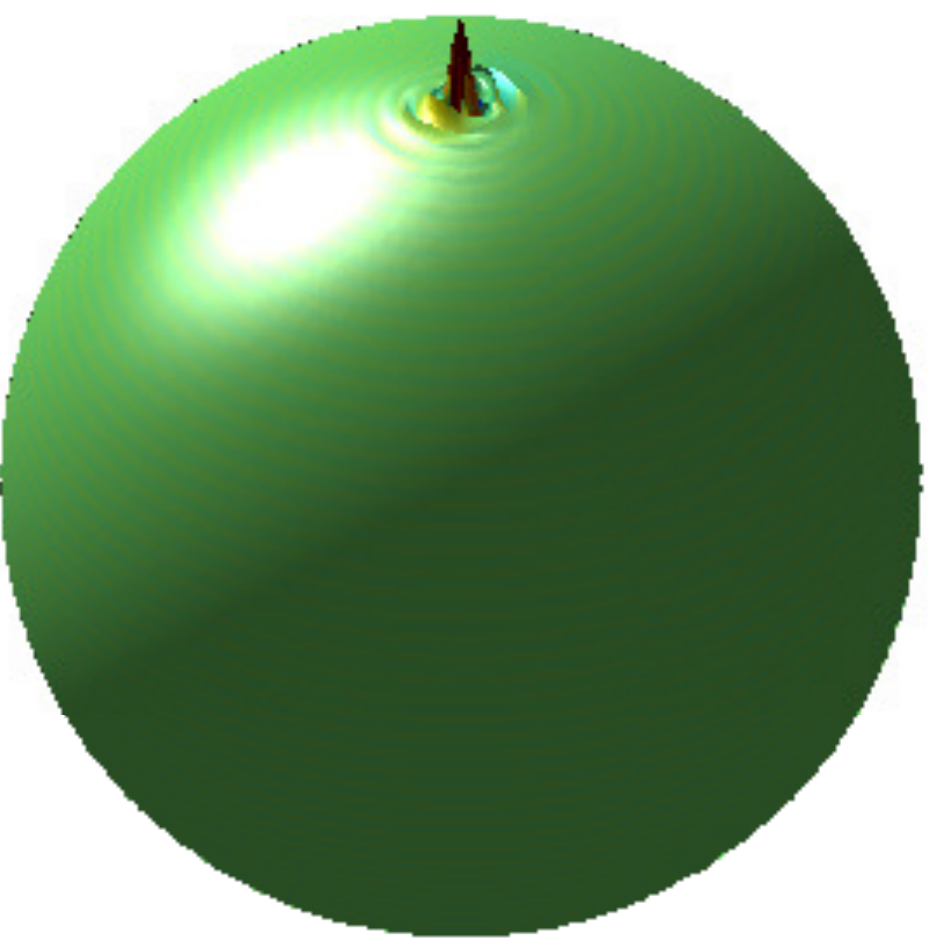}}\hfil
\subfigure[$\wscale=1$]{\includegraphics[width=0.28\textwidth]{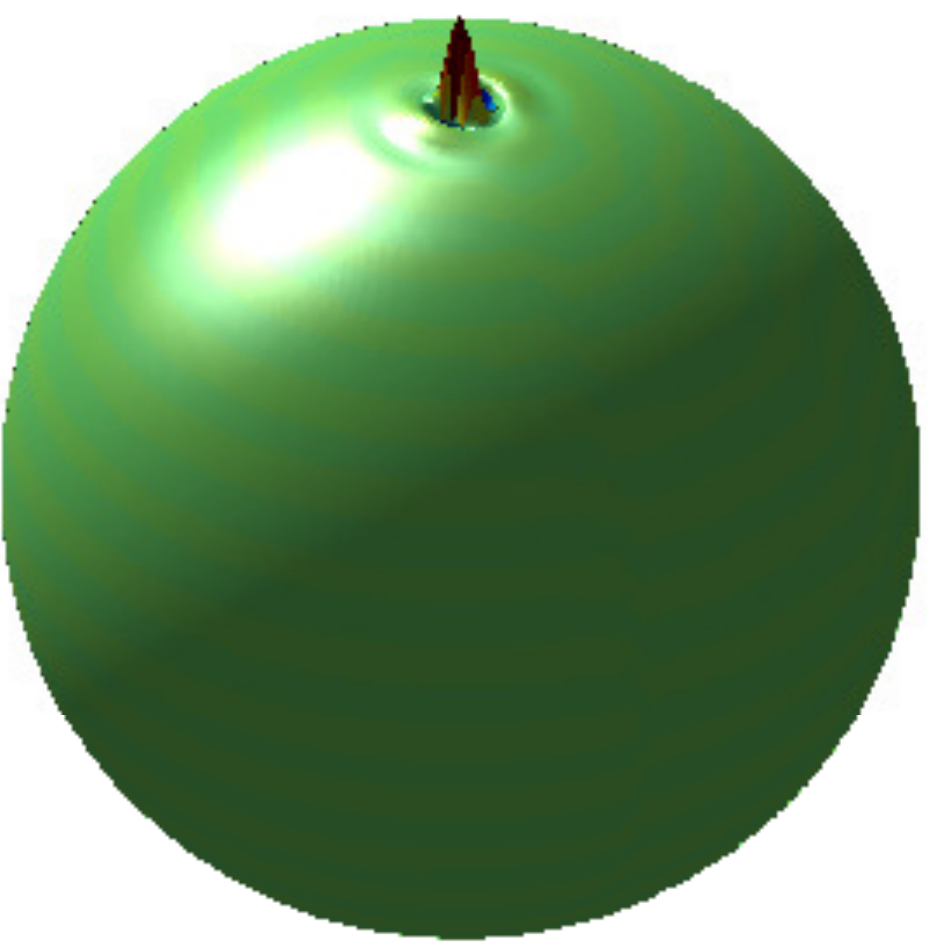}}\hfil
\subfigure[$\wscale=2$]{\includegraphics[width=0.28\textwidth]{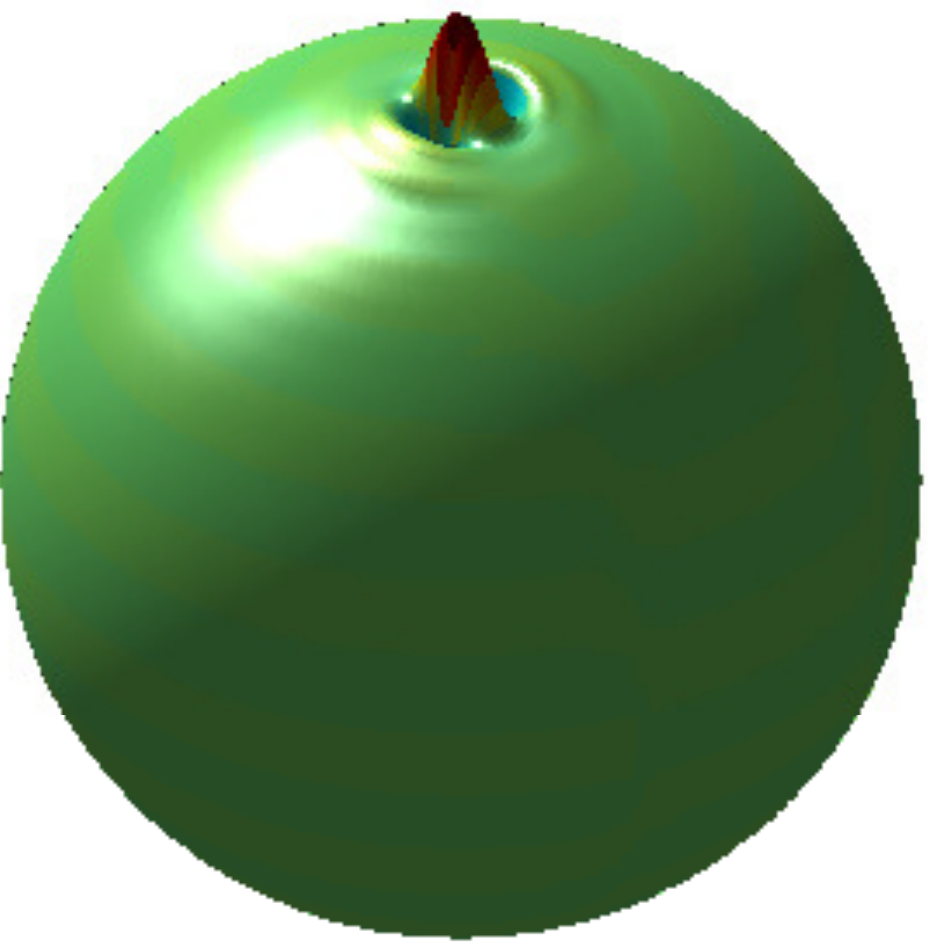}} \\
\subfigure[$\wscale=3$]{\includegraphics[width=0.28\textwidth]{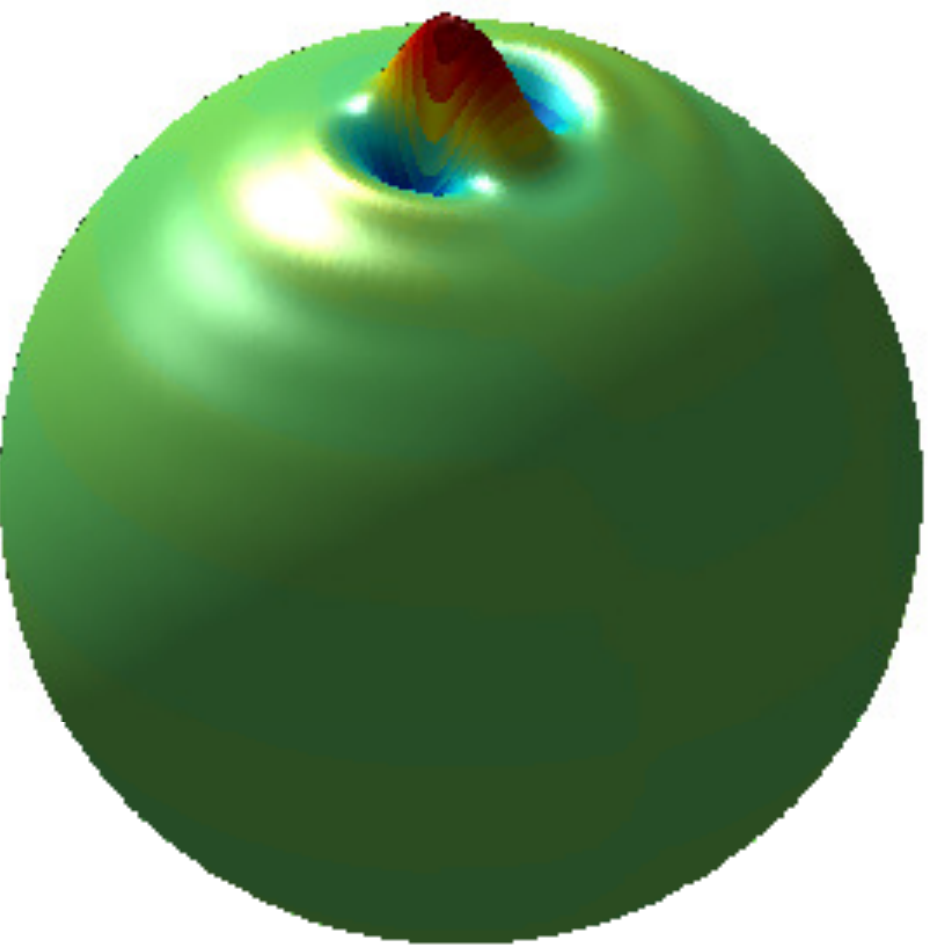}}\hfil
\subfigure[$\wscale=4$]{\includegraphics[width=0.28\textwidth]{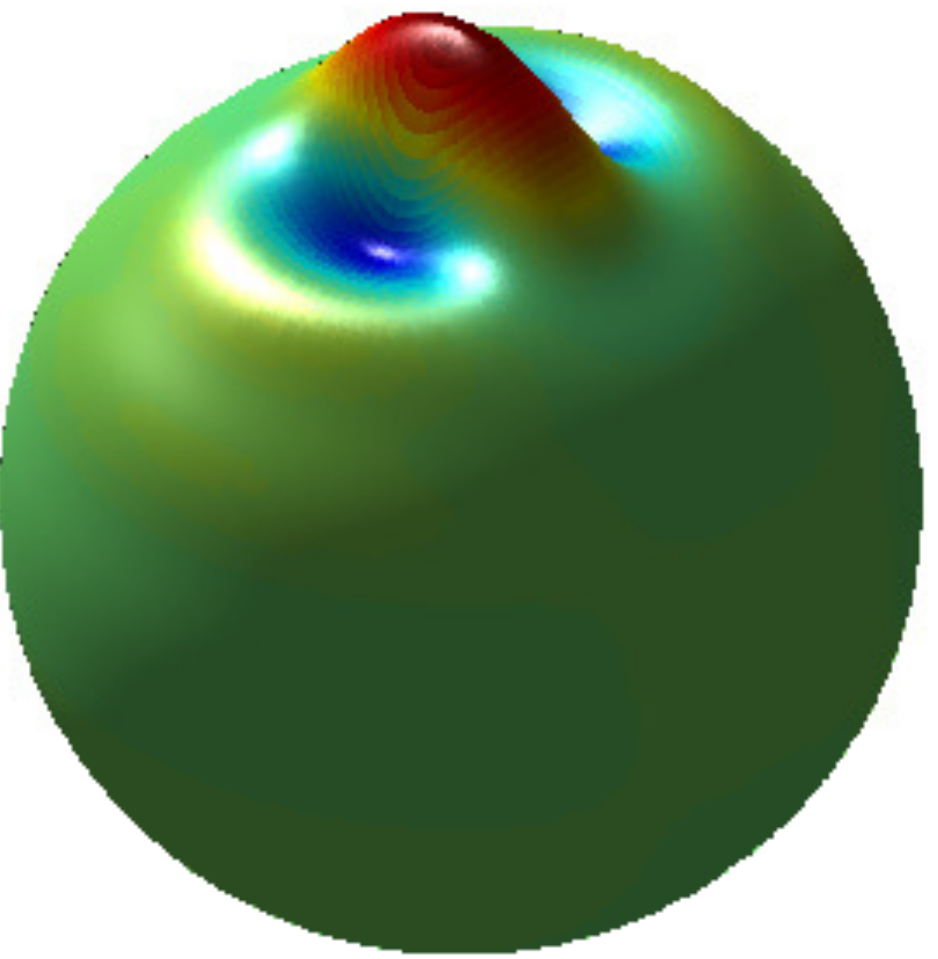}}\hfil
\subfigure[$\wscale=5$]{\includegraphics[width=0.28\textwidth]{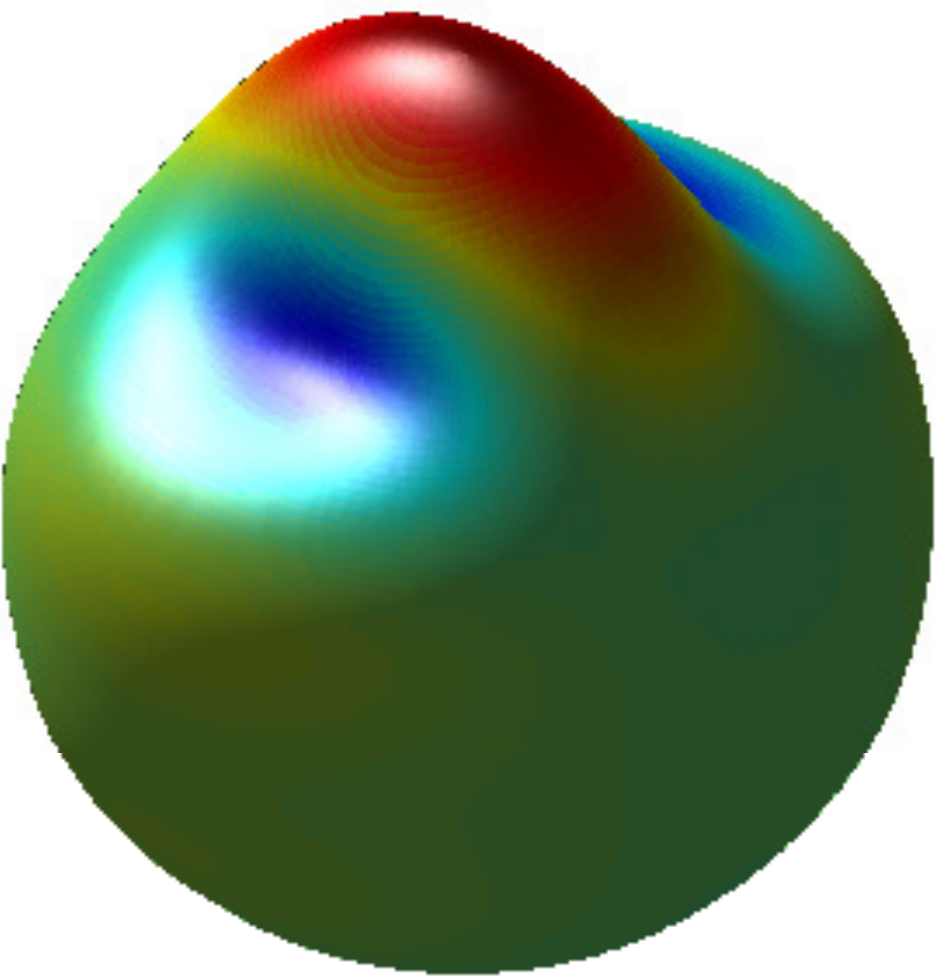}} 
\caption{Parametric plots of scale-discretized wavelets on the sphere ($\elmax=128$, $\nmax=3$,
  $\wscalemax=5$, $\dilparam=2$).}
\label{fig:wavelets}
\end{figure}

For the wavelets, scaling function and scale parameter ranges outlined
above, the admissibility criterion \eqn{\ref{eqn:admissibility}} is
satisfied.  Example directional scale-discretized wavelets are plotted
in \fig{\ref{fig:wavelets}}.

\section{EXACT AND EFFICIENT COMPUTATION} 
\label{sec:computation}

We describe in this section the algorithms implemented in the {\tt
  S2DW}\footnote{\url{http://www.s2dw.org/}} package to support the
exact and efficient computation of the scale-discretized wavelet
transform on the sphere.  We focus on the case of band-limited
signals, which admit exact quadrature rules for certain
discretizations of the sphere \sphere\ and rotation group \sothree,
leading to the exact computation of wavelet analysis and synthesis,
\ie\ forward and inverse transforms, respectively.  Furthermore, we
exploit the factoring of rotations approach to develop fast algorithms
for both analysis and synthesis.  Finally, we perform numerical
experiments to illustrate the speed and accuracy of the algorithms
developed.

\subsection{Discretization and quadrature} 
\label{sec:computation:quadrature}

The Driscoll and Healy sampling theorem\cite{driscoll:1994,healy:2003}
is adopted in {\tt S2DW}, with the corresponding cubature points:
\mbox{$\saaiang= \pi(2\saai+1) / (4\elmax) $}, for $\saai = 0, \dotsc,
2\elmax-1$, and $\sabiang = 2\pi\sabi / (2\elmax-1)$, for $\sabi = 0,
\dotsc, 2\elmax -2$, giving $4\elmax^2$ samples on the sphere.
The Driscoll and Healy sampling theorem can be distilled into the
following quadrature rule for the \emph{exact} integration over
colatitude \saa\ of a function \f\ band-limited at $2\elmax$ or below:
\begin{equation}
  \label{eqn:quadrature_dh}
  \int_0^\pi \dx \saa \sin\saa \: \f(\saa,\cdot)
  = \sum_{\saai=0}^{2\elmax-1} \qweight(\saaiang) \f(\saaiang,\cdot)
  \spcend,
\end{equation}
where the quadrature weights are given
by\cite{driscoll:1994,healy:2003,mcewen:2011:waveletsxiv} 
\begin{equation}
  \qweight(\saaiang) =
  \frac{2}{\elmax} \:
  \sin\saaiang \:
  \sum_{k=0}^{\elmax-1} \:
  \frac{\sin((2k+1)\saaiang)}{2k+1}
  \spcend .
\end{equation}
We will also make use of exact quadrature for the following
integration over longitude \sab, for a function \f\ band-limited
at $\elmax$:
\begin{equation}
  \label{eqn:quadrature_ft}
  \int_0^{2\pi} \dx \sab \: {\rm exp}(-i\m\sab) \: \f(\cdot,\sab)
  = \frac{2\pi}{2\elmax-1} 
  \sum_{\sabi=0}^{2\elmax-2}   {\rm exp}(-i\m\sabiang) \f(\cdot,\sabiang)
  \spcend,
\end{equation}
which follows from the continuous and discrete orthogonality of the
complex exponentials.

We have so far considered the discretization of functions defined on
the sphere \sphere; however, the wavelet coefficients themselves are
defined on the rotation group \sothree\ (due to the directional nature
of the wavelet transform).  The Euler angles $\eul=(\euls)$ that
parameterize \sothree\ can be discretized by making the association to
the sphere $(\eulbiang,\eulaiang) = (\saaiang,\sabiang)$, where the
indices $\eulai$ and $\eulbi$ vary over the same range as $\sabi$ and
$\saai$ respectively, and by discretizing \eulc\ by $\eulciang = \pi
\eulci / \nmax$, for $\eulci = 0,\ldots,\nmax-1$.  Note that the
wavelets constructed in \sectn{\ref{sec:transform:construction}} are
invariant under an azimuthal rotation by $\pi$ (when centered on the
North pole), hence it is only necessary to discretize \eulc\ in the
range $[0,\pi)$.

Finally, we note that in order to recover exactness for the wavelet
analysis and synthesis algorithms that follow, we assume that the
spherical harmonic coefficients of the function of interest can be
computed exactly by appealing to a sampling theorem on the sphere
(\eg\ the Driscoll \& Healy sampling
theorem\cite{driscoll:1994,healy:2003}).  In this setting all
algorithms are theoretically exact.  If a sampling theorem is not
available for the particular pixelization of the sphere on which
samples are taken (\eg\ if one adopts the \healpix\ pixelization
scheme\cite{gorski:2005}), harmonic coefficients can nevertheless be
computed approximately and the algorithms presented below applied.

\subsection{Wavelet analysis} 
\label{sec:computation:analysis}

The wavelet analysis given by \eqn{\ref{eqn:analysis}} may be
expressed in harmonic space by
\begin{equation}
  \label{eqn:analysis_harmonic}
  \wcoeff^{\wav^\wscale}(\eul) 
  = \sumlmn
  \shc{\f}{\el}{\m}
  \shc{\wav}{\el}{\n}^{\wscale\cconj}
  \Dlmnpc  
\end{equation}
where the Wigner $\dmatbig$-functions \Dlmn\ are the matrix elements
of the irreducible unitary representation of the rotation group
\sothree.  Consequently, the \Dlmnc\ also form an orthogonal basis in
$\ltwo(\sothree)$.  Since we consider functions band-limited at
\elmax\ and directional wavelets with azimuthal band-limit $\nmax$ in
what follows, the upper limits of the summation over \el\ in
\eqn{\ref{eqn:analysis_harmonic}} can be truncated to $\elmax-1$ and
the upper (lower) limit in the summation over \n\ can be truncated to
$\nmax-1$ ($-\nmax+1$).  Subsequently, we use the shorthand notation
$\sumulmn$ to represent this case. In general, for the sake of brevity
and readability we do not specify the limits of summation in the
following, since the limits can be inferred easily.  Since the
computation of wavelet coefficients is expressed in
\eqn{\ref{eqn:analysis_harmonic}} as a finite sum rather than an
integral (thanks for the orthogonality of the spherical harmonics),
the wavelet coefficients may be computed exactly.

Wavelet analysis can be seen as an inverse Wigner transform, where the
Wigner coefficients corresponding to the wavelet coefficients are given
by
\begin{equation}
   \wigc{\bigl(\wcoeff^{\wav^\wscale}\bigr)}{\el}{\m}{\n} 
   = 
   \frac{8\pi^2}{2\el+1}
   \shc{\f}{\el}{\m}
   \shc{\wav}{\el}{\n}^{\wscale\cconj}
   \spcend.
\end{equation}
Note that we have adopted the following convention for the Wigner
transform: a function $g \in \ltwo(\sothree)$ may be expanded in terms
of the basis functions \Dlmnc\ by
\begin{equation}
   g(\eul) = \sum_\el \frac{2\el+1}{8\pi^2} \sum_{\m\n} 
   \wigc{g}{\el}{\m}{\n} \Dlmnpc
   \spcend,
\end{equation}
where the Wigner coefficients are given by
\begin{equation}
  \wigc{g}{\el}{\m}{\n} = \innerp{g}{\Dlmnc} 
  = \int_\sothree \deul{\eul} g(\eul) \Dlmnp
  \spcend.
\end{equation}

Wavelet analysis can be computed efficiently using the factoring of
rotations approach\cite{risbo:1996,wandelt:2001,mcewen:2006:fcswt} as
first suggest and computed in Ref.~\citenum{mcewen:2006:fcswt}.
We note the $\dmatbig$-function decomposition in terms of the real
$\dmatsmall$-functions:\cite{varshalovich:1989}
\begin{equation}
  \label{eqn:wigner_decomposition}
  \Dlmn(\euls) = 
  {\rm exp}(-i\m\eula) \:
  \dlmn(\eulb) \:
  {\rm exp}(-i\n\eulc)
  \spcend ,
\end{equation}
and the Fourier expansion of the \dlmn:
\begin{equation}
  \label{eqn:wigner_fourier_decomposition}
  \dlmn(\eulb) 
  = 
  {\rm exp}(-i(\n-\m)\pi/2)
  \sum_{\m\p}
  \dlmnhalfpi{\el}{\m\p}{\m}
  \dlmnhalfpi{\el}{\m\p}{\n}
  {\rm exp}(-i\m\p\eulb)
  \spcend ,
\end{equation}
which follows by a factoring of
rotations\cite{risbo:1996,wandelt:2001,mcewen:2006:fcswt}, where $
\dlmnhalfpi{\el}{\m}{\n} = \dlmn(\pi/2)$.  Substituting these
expressions into \eqn{\ref{eqn:analysis_harmonic}} and interchanging
the order summation, one finds:
\begin{equation}
  \label{eqn:analysis_fast1}
  \wcoeff^{\wav^\wscale}(\eul) 
  = \sum_{\m\m\p\n}
  U_{\m\m\p\n} \:
  {\rm exp}(i(\m\eula + \m\p\eulb + \n\eulc))
  \spcend ,
\end{equation}
where 
\begin{equation}
  \label{eqn:analysis_fast2}
  U_{\m\m\p\n} 
  = {\rm exp}(i(\n-\m)\pi/2)
  \sum_\el 
  \dlmnhalfpi{\el}{\m\p}{\m}
  \dlmnhalfpi{\el}{\m\p}{\n}
  \shc{\f}{\el}{\m}
  \shc{\wav}{\el}{\n}^{\wscale\cconj}
  \spcend .
\end{equation}

The first step of the fast wavelet analysis algorithm is to compute
$U_{\m\m\p\n}$ via \eqn{\ref{eqn:analysis_fast2}}, which can be
computed with asymptotic complexity $\order(\nmax \elmax^3)$.  The next
step is to compute wavelet coefficients from $U_{\m\m\p\n}$ via
\eqn{\ref{eqn:analysis_fast1}}. In Ref.~\citenum{mcewen:2006:fcswt} a
three-dimensional fast Fourier transform (FFT) was applied to perform
this computation efficiently.  However, in that setting the
discretization of the Euler angles $(\euls)$ was largely arbitrary
since wavelet synthesis was not considered (indeed, not possible in
practice due to the continuous nature of the wavelet transform
considered).  Here we must evaluate the wavelet coefficients at the
sample positions for which we have exact quadrature rules, as outlined
in \sectn{\ref{sec:computation:quadrature}}.  We thus consider each
Euler angle in turn and perform a separation of variables to compute 
\eqn{\ref{eqn:analysis_fast1}} efficiently.

Firstly, consider the Euler angle \eulc:
\begin{equation}
  U_{\m\m\p}(\eulc) = 
  \sum_\n  U_{\m\m\p\n} \: {\rm exp}(i\n\eulc)
  \spcend .
\end{equation}
Since the domain of interest for \eulc\ is $[0,\pi)$ and the azimuthal
band-limit \nmax\ is often very low, \ie\ $\nmax \ll \elmax$, this
summation is computed explicitly, with complexity
$\order(\nmax^2 \elmax^2)$. 

Next, consider the Euler angle \eulb:
\begin{equation}
  U_{\m}(\eulb,\eulc) = 
  \sum_{\m\p}  U_{\m\m\p}(\eulc) \: {\rm exp}(i\m\p\eulb)
  \spcend .
\end{equation}
Since we discretize \eulb\ at the sample positions supported by the
Driscoll and Healy sampling theorem, as specified in
\sectn{\ref{sec:computation:quadrature}}, this summation is also
computed explicitly, with complexity $\order(\nmax \elmax^3)$.

Finally, consider the Euler angle \eula:
\begin{equation}
  \wcoeff^{\wav^\wscale}(\eul)  = 
  \sum_{\m}  U_{\m}(\eulb,\eulc) \: {\rm exp}(i\m\eula)
  \spcend.
\end{equation}
Since we are free to choose the discretization of \eula\ we ensure our
choice is compatible with an FFT, so that the complexity of this
computation is reduced from $\order(\nmax \elmax^3)$ for a direct
computation to $\order(\nmax \elmax^2 \log\elmax)$ by application of
an FFT. 

To summarise, wavelet analysis given by \eqn{\ref{eqn:analysis}} can
be computed exactly by its representation as a discrete sum in harmonic
space.  Furthermore, we have described a fast algorithm for the exact
computation of wavelet coefficients by performing a factoring of
rotations and separation of variables.  This fast algorithm reduces the
complexity of computing wavelet coefficients for a given scale
\wscale\ from $\order(\elmax^5)$ to $\order(\nmax \elmax^3)$.  Since
$\wscalemax+1$ wavelet scales are considered, the overall complexity
of computing all wavelet coefficients is $\order(\wscalemax \nmax
\elmax^3 )$.  However, since the wavelets themselves have compact
support it is not necessary to perform all wavelet transforms at the
full band-limit \elmax.  Furthermore, the lower support of the
wavelets can also be exploited to increase computational efficiency.
Following these optimizations the computation is dominated by the
largest two scales.  Hence, the overall complexity of computing all
wavelet transforms for all scales is effectively $\order(\nmax
\elmax^3)$. 

We close this discussion of the computation of the wavelet transform by noting that
scaling coefficients can also be computed exactly and efficiently.
Since the scaling coefficients are given by an axisymmetric
convolution via \eqn{\ref{eqn:analysis_scaling}}, their harmonic
coefficients are given by
\begin{equation}
  \shc{\wcoeff}{\el}{\m}^\wavs
  = \sqrt{\frac{4\pi}{2\el+1}}
  \shc{\f}{\el}{\m}
  \shcc{\wavs}{\el}{0}
  \spcend ,
\end{equation}
which can be computed in $\order(\elmax^2)$. The scaling coefficients
can then be computed exactly in real space by appealing to a sampling
theorem on the sphere.

\subsection{Wavelet synthesis} 
\label{sec:computation:synthesis}

We first consider the contribution to the wavelet synthesis expression
of \eqn{\ref{eqn:synthesis}} from the wavelet coefficients.
Expressing this contribution in harmonic space one finds:
\begin{equation}
  \label{eqn:synthesis_fast1}
  \sum_{\wscale=0}^\wscalemax \int_\sothree \deul{\eul}
  \wcoeff^{\wav^\wscale}(\eul) (\rotarg{\eul} L^{\rm d} \wav^\wscale)(\sa)
  = \sum_{\wscale=0}^\wscalemax \sum_{\el\m\n}
  \frac{2\el+1}{8\pi^2} 
  \wigc{\bigl(\wcoeff^{\wav^\wscale}\bigr)}{\el}{\m}{\n} 
  \shc{\wav}{\el}{\n}^\wscale
  \shfarg{\el}{\m}{\sa}
  \spcend,
\end{equation}
where
\begin{equation}
  \label{eqn:synthesis_fast2}
  \wigc{\bigl(\wcoeff^{\wav^\wscale}\bigr)}{\el}{\m}{\n} 
  = \innerp{\wcoeff^{\wav^\wscale}}{\Dlmnc} 
  = \int_\sothree \deul{\eul} \wcoeff^{\wav^\wscale} (\eul) \Dlmnp
  \spcend.
\end{equation}
Thus, the wavelet coefficient contribution to the function synthesis
can be computed exactly by \eqn{\ref{eqn:synthesis_fast1}}, with
complexity $\order(\nmax \elmax^2)$ for each $\wscale$ (we consider
all \wscale\ at the end of this subsection) to compute
the contribution to the harmonic coefficients $\shc{\f}{\el}{\m}$,
provided the Wigner coefficients
$\wigc{\bigl(\wcoeff^{\wav^\wscale}\bigr)}{\el}{\m}{\n}$ can be
computed exactly.  We therefore turn our attention to the exact and
efficient computation of \eqn{\ref{eqn:synthesis_fast2}}.

\eqn{\ref{eqn:synthesis_fast2}} can also be computed efficiently using
the factoring of rotations approach and a separation of variables.  Substituting the
$\dmatbig$-function decomposition
\eqn{\ref{eqn:wigner_decomposition}} and the Fourier expansion of
$\dlmn$ \eqn{\ref{eqn:wigner_fourier_decomposition}},
\eqn{\ref{eqn:synthesis_fast2}} reads
\begin{equation}
  \label{eqn:synthesis_fast3}
  \wigc{\bigl(\wcoeff^{\wav^\wscale}\bigr)}{\el}{\m}{\n} 
  =
  {\rm exp}(-i(\n-\m)\pi/2)
  \sum_{\m\p} 
  \dlmnhalfpi{\el}{\m\p}{\m}
  \dlmnhalfpi{\el}{\m\p}{\n}
  V_{\m\m\p\n}
  \spcend,
\end{equation}
where
\begin{equation}
  \label{eqn:synthesis_fast4}
  V_{\m\m\p\n} 
  =
  \int_\sothree \deul{\eul}
  \wcoeff^{\wav^\wscale}(\euls) \:
  {\rm exp}(-i(\m\eula + \m\p\eulb + \n\eulc))
  \spcend.
\end{equation}
Note that \eqn{\ref{eqn:synthesis_fast3}} can be computed exactly with
complexity $\order(\nmax \elmax^3)$.  We next appeal to the quadrature
rules described in \sectn{\ref{sec:computation:quadrature}} to
evaluate \eqn{\ref{eqn:synthesis_fast4}} exactly, considering each
Euler angle in turn.

Firstly, consider the Euler angle \eula.  Appealing to the quadrature
rule \eqn{\ref{eqn:quadrature_ft}}, \eqn{\ref{eqn:synthesis_fast4}}
may be written:
\begin{equation}
  \label{eqn:synthesis_fast5}
  V_{\m\m\p\n}
  =
  \int_0^{\pi} \dx\eulb \sin\eulb
  \int_0^{2\pi} \dx\eulc \:
  V_{\m}(\eulb,\eulc) \:
  {\rm exp}(-i(\m\p\eulb + \n\eulc))
  \spcend,
\end{equation}
where
\begin{equation}
  \label{eqn:synthesis_fast6}
  V_{\m}(\eulb,\eulc)
  =
  \frac{2\pi}{2\elmax-1}
  \sum_\eulai
  \wcoeff^{\wav^\wscale}(\eulaiang,\eulb,\eulc) \:
  {\rm exp}(-i \m\eulaiang )
  \spcend.
\end{equation}
The Euler angle \eula\ is discretized according to the scheme
specified in \sectn{\ref{sec:computation:quadrature}}, where $\eulai$
denotes the index associated with $\eula$.  Naively
\eqn{\ref{eqn:synthesis_fast6}} may be computed with $\order(\nmax
\elmax^3)$; however, we exploit an FFT to reduce this to $\order(\nmax
\elmax^2 \log\elmax)$.

Next, consider the Euler angle \eulc.  We exploit the steerability
of wavelet coefficients \eqn{\ref{eqn:steerability_wcoeff}}, which transfers
  to $V_{\m}(\eulb,\eulc)$ directly, to compute integration over \eulc\
  exactly and efficiently:
\begin{equation}
  \label{eqn:synthesis_fast7}
  V_{\m\m\p\n}
  =
  \int_0^{\pi} \dx\eulb \sin\eulb \:
  V_{\m\n}(\eulb) \:
  {\rm exp}(-i\m\p\eulb)
  \spcend,
\end{equation}
where
\begin{equation}
  \label{eqn:synthesis_fast8}
  V_{\m\n}(\eulb)
  =
  2 \pi z_\n
  \sum_\eulci
  V_{\m}(\eulb,\eulciang) \:
  {\rm exp}(-i \n\eulciang )
  \spcend,
\end{equation}
which can be computed in $\order(\nmax^2 \elmax^2)$.  The Euler angle
\eulc\ is discretized according to the scheme specified in
\sectn{\ref{sec:computation:quadrature}}, where $\eulci$ denotes the
index associated with $\eulc$.

Finally, consider the Euler angle \eulb.  Appealing to the quadrature
rule \eqn{\ref{eqn:quadrature_dh}}, which follows from the Driscoll
and Healy sampling theorem, \eqn{\ref{eqn:synthesis_fast7}} may be
written:
\begin{equation}
  \label{eqn:synthesis_fast9}
  V_{\m\m\p\n}
  =
  \sum_\eulbi
  \qweight(\eulbiang)
  V_{\m\n}(\eulbiang) \:
  {\rm exp}(-i \m\p\eulbiang )
  \spcend,
\end{equation}
which can be computed in $\order(\nmax \elmax^3)$.  The Euler angle
\eulb\ is discretized according to the scheme specified in
\sectn{\ref{sec:computation:quadrature}}, where $\eulbi$ denotes the
index associated with $\eulb$.

To summarise, the wavelet coefficient contribution to the function
synthesis can be computed exactly by appealing to the quadrature rules
outlined in \sectn{\ref{sec:computation:quadrature}}.  Furthermore, we
have described a fast algorithm for this exact computation based on a
factoring of rotations and separation of variables.  This fast
algorithm reduces the complexity of the computation from
$\order(\elmax^5)$ for the naive case to $\order(\nmax \elmax^3)$.
Since $\wscalemax+1$ wavelet scales are considered the overall
complexity becomes $\order(\wscalemax \nmax \elmax^3)$.  However,
again the compact support of the wavelets may be exploited so that the
computation is dominated by the largest two scales, resulting in an
overall complexity of effectively $\order(\nmax \elmax^3)$.

We finally consider the contribution to the function synthesis
expression of \eqn{\ref{eqn:synthesis}} from the scaling
coefficients.  Expressing this contribution in harmonic space one
finds: 
\begin{equation}
  2\pi \int_\sphere \dmu{\sa\p} 
  \scoeff^\wavs(\sa\p) (\rotarg{\sa\p} L^{\rm d} \wavs)(\sa)
  =
  \sum_{\el\m} \sqrt{\frac{2\el+1}{4\pi}}
  \shc{\wcoeff}{\el}{\m}^\wavs
  \shc{\wavs}{\el}{0}
  \shfarg{\el}{\m}{\sa}
  \spcend .
\end{equation}
The scaling coefficient contribution to the harmonic coefficients of
the synthesized signal can thus be computed in $\order(\elmax^2)$.

\subsection{Numerical experiments} 

The exact and efficient algorithms described above to compute wavelet
analysis and synthesis, \ie\ forward and inverse wavelet transforms
respectively, are implemented in the {\tt
  S2DW}\footnote{\url{http://www.s2dw.org/}} package, which is
publicly available.  The {\tt S2DW} package is written in {\tt
  Fortran}, uses the {\tt FFTW}\footnote{\url{http://www.fftw.org/}}
to perform Fourier transforms, and (in version 1.1) is parallelized to
run on multi-core architectures.\footnote{The alternative {\tt
    S2LET}\cite{leistedt:s2let_axisym} package
  (\url{http://www.s2let.org/}) provides a C implementation of
  scale-discretized wavelets, with Matlab interfaces.  However, at
  present {\tt S2LET} supports axisymmetric wavelets only.}  Here we
evaluate the accuracy and efficiency of {\tt S2DW} up to extremely
high band-limits ($\elmax=4096$), corresponding to functions sampled
at tens of millions of pixels on the sphere.  For reference, this
band-limit exceeds the resolution of the recently released
\emph{Planck}\cite{planck2013-p01} full-sky observations of the \cmb.

We perform the following numerical experiments.  For a given
band-limit \elmax, we generate random test signals \f\ with harmonic
coefficients uniformly randomly distributed in $[-1,1]$.  We compute
all wavelet coefficients (up to the maximum possible wavelet scale
$\wscalemax=\wscalemax_\elmax(\dilparam)$) by applying the wavelet
analysis algorithm detailed in \sectn{\ref{sec:computation:analysis}}.
We then synthesis the function from its wavelet coefficients by
applying the wavelet synthesis algorithm detailed in
\sectn{\ref{sec:computation:synthesis}}.  Numerical accuracy is
evaluated by the maximum error between the original harmonic
coefficients of \f, denoted $\shc{\f}{\el}{\m}^{\rm orig}$, and their
synthesised values $\shc{\f}{\el}{\m}^{\rm recon}$: $\epsilon =
\max_{\el,\m} \vert\shc{\f}{\el}{\m}^{\rm recon} -
\shc{\f}{\el}{\m}^{\rm orig} \vert$. Furthermore, we also record the
wall-time clock $\tau$ for the combined analysis-synthesis
computation.  These numerical experiments are performed on a 2x Intel
Xeon X5650 Processor (2.66Ghz, 12M Cache, 6.40 GT/s QPI, Turbo, HT)
machine with 12 cores, where we perform 10 simulations for each
band-limit considered.

\begin{figure}
\centering
\subfigure[Accuracy\label{fig:performance:accuracy}]{\includegraphics[width=0.48\textwidth]{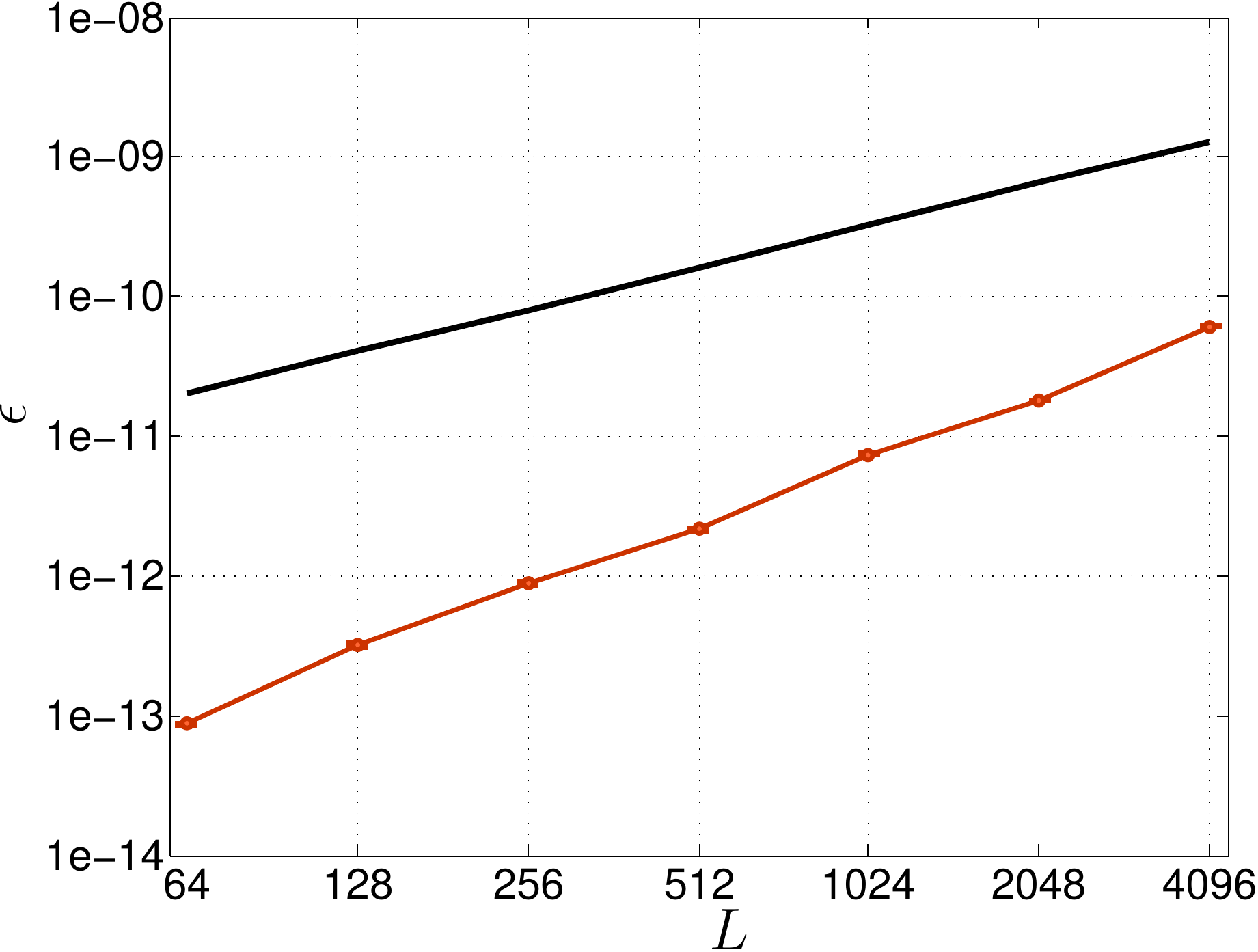}}
\quad
\subfigure[Timing\label{fig:performance:timing}]{\includegraphics[width=0.48\textwidth]{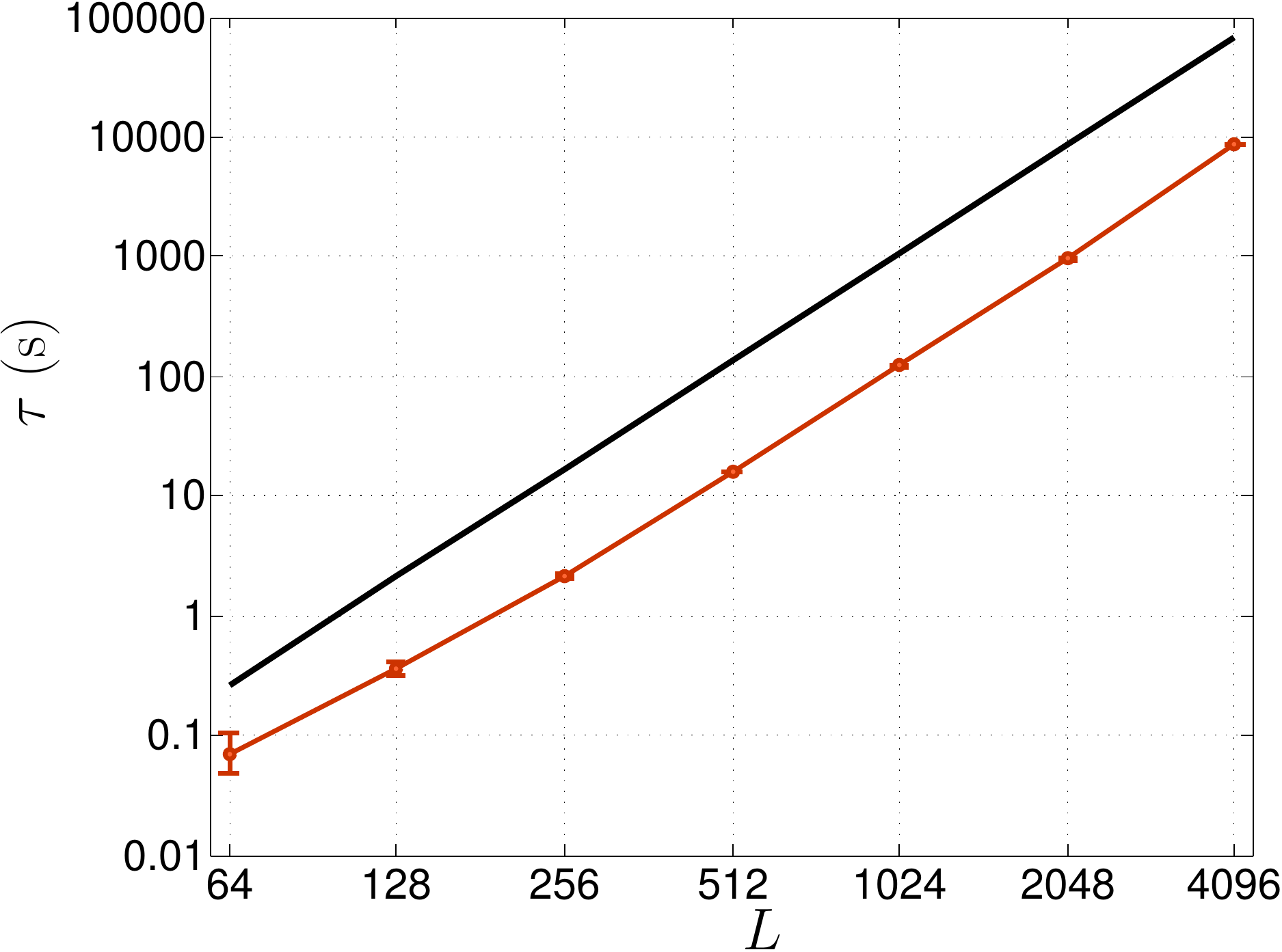}}
\caption{Computational performance of the {\tt S2DW} implementation of
  the scale-discretized wavelet transform on the sphere ($\nmax=3$,
  $\wscalemax=\wscalemax_\elmax(\dilparam)$, $\dilparam=2$).  In
  panel~(a) the numerical error following a forward and inverse
  wavelet transform is shown.  The algorithms described herein are
  theoretically exact and achieve close to machine precision for
  extremely large band-limits.  Numerical error is found empirically
  to scale as approximately $\order(\elmax)$, illustrated by the solid
  curve.  In panel~(b) the computation time of a forward and inverse
  wavelet transform is shown.  As predicted, computation time is shown
  to scale as $\order(\elmax^3)$, illustrated by the solid curve.  In
  both panels the mean of 10 simulations is shown, with one standard
  deviation error bars.  In most cases the errors bars are too small
  to be seen clearly.}
\label{fig:performance}
\end{figure}

The results of these numerical experiments are shown in
\fig{\ref{fig:performance}}.  Firstly, notice that very good numerical
accuracy is achieved (\fig{\ref{fig:performance:accuracy}}), at the
level of machine precision, up to very high band-limits.  The
numerical error is found empirically to scale as approximately
$\order(\elmax)$.  Secondly, the algorithms discussed above are
demonstrated to scale as $\order(\elmax^3)$
(\fig{\ref{fig:performance:timing}}) as predicted.  Furthermore, the
algorithms are sufficiently fast to allow computation up to very high
band-limits in a reasonable time.  These computation times are for
version 1.1 of {\tt S2DW} and are a considerable improvement over the
computation time specified in Ref.~\citenum{wiaux:2007:sdw} for
version 1.0, due to parallelization, other code optimizations, and
advances in computing hardware.  For example, the round-trip
computation time for $\elmax=1024$ is reduced from 72 minutes to less
than 2 minutes.

\section{SUMMARY AND FUTURE PERSPECTIVES} 
\label{sec:summary}

We have described exact and efficient algorithms to compute the
wavelet analysis and synthesis of the scale-discretized wavelet
transform on the sphere developed by Wiaux \etal\cite{wiaux:2007:sdw}.
These algorithms rely on the factoring of rotations
approach\cite{risbo:1996,wandelt:2001,mcewen:2006:fcswt} and a
separation of variables.  The compact support of the wavelets and
azimuthal band-limit \nmax\ may be exploited to further improve
computational efficiency.  Consequently, the cost of computing wavelet
transforms is reduced from $\order(\wscalemax \elmax^5)$ for the naive
setting to effectively $\order(\nmax \elmax^3)$, for both analysis and
synthesis algorithms, \ie\ for both the forward and inverse wavelet
transforms.

The exactness of the algorithms is achieved by appealing to the
Driscoll and Healy sampling theorem\cite{driscoll:1994,healy:2003} on
the sphere.  However, there is scope to further optimize this part of
the algorithms, \ie\ integrals and summations over \saa\ and \eulb.
Recently, two of the authors of this article developed a novel
sampling theorem on the sphere,\cite{mcewen:fssht} superseding the
Driscoll and Healy sampling theorem.  This new sampling theorem
reduces the number of samples on the sphere required to represent a
band-limited signal from $\sim 4 \elmax^2$ to $\sim 2 \elmax^2$.
Furthermore, FFTs are exploited to yield fast algorithms to compute
spherical harmonic transforms associated with the new sampling
theorem rapidly.\cite{mcewen:fssht} In future work we plan to integrate these
recent developments into the algorithms presented herein, replacing
the use of the Driscoll and Healy sampling theorem with the new
sampling theorem of Ref.~\citenum{mcewen:fssht}.  This will reduce the
number of sample values required to represent wavelet coefficients
(and functions on the sphere) by a factor of two, while retaining all
of their information content, and will further improve the speed of
the algorithms to compute the scale-discretized wavelet transform on
the sphere.


\acknowledgments     
 
JDM is supported in part by a Newton International Fellowship from the
Royal Society and the British Academy. YW is supported in part by the
Center for Biomedical Imaging (CIBM) of the Geneva and Lausanne
Universities, EPFL and the Leenaards and Louis-Jeantet foundations.
The numerical experiments presented in this article were carried out
using facilities funded by STFC and Marie Curie grant
MIRG-CT-2007-203314 from the European Commission.


\bibliography{bib}   
\bibliographystyle{spiebib}   

\end{document}